# A 'p-n' diode with Hole and Electron-doped lanthanum manganites

C. Mitra[a1], P. Raychaudhuri[b2], G. Köbernik[1], K. Dörr[1], K-H. Müller[1], R. Pinto[3]

[1] Institut für Festkörper- und Werkstofforschung Dresden,
Helmholtzstraße 20, D-01069 Dresden, Germany

[2] School of Physics and Astronomy,
University of Birmingham,
Edgbaston, Birmingham,
B15 2TT, UK.

[3] Department of Condensed Matter Physics & Material Science,
Tata Institute of Fundamental Research, Homi Bhabha Road,
Colaba, Mumbai 400 005, India.

## Abstract

The hole-doped manganite $La_{0.7}Ca_{0.3}MnO_3$ and the electron-doped manganite $La_{0.7}Ce_{0.3}MnO_3$ undergo an insulator to metal transition at around 250 K, above which both behave as a polaronic semiconductor. We have successfully fabricated an epitaxial trilayer ($La_{0.7}Ca_{0.3}MnO_3/SrTiO_3/La_{0.7}Ce_{0.3}MnO_3$), where $SrTiO_3$ is an insulator. At room temperature, i.e. in the semiconducting regime, it exhibits asymmetric current-voltage (I-V) characteristics akin to a p-n diode. The observed asymmetry in the I-V characteristics disappears at low temperatures where both the manganite layers are metallic. To the best of our knowledge, this is the first report of such a p-n diode, using the polaronic semiconducting regime of doped manganites.

[a] e-mail: chiranjib@ifw-dresden.de

[b] e-mail: P.Raychaudhuri@bham.ac.uk

LaMnO$_3$ is an antiferromagnetic insulator, with the Mn ions existing in the Mn$^{3+}$ valence state, its electronic configuration being $t_{2g}^3 e_g^1$. When it is hole-doped by replacing La$^{3+}$ ions partially by Ca$^{2+}$ ions, the charge state of an equivalent number of Mn ions changes to Mn$^{4+}$, thus permitting a hopping of $e_g$ electrons between spin aligned Mn$^{3+}$ and Mn$^{4+}$ ions giving rise to a double exchange interaction [1-3], which results in an effective ferromagnetic interaction between the Mn$^{3+}$ and Mn$^{4+}$ ions. The compound La$_{0.7}$Ca$_{0.3}$MnO$_3$ has an ordering temperature (T$_c$) of around 250 K and undergoes a "semiconductor-metal" transition below T$_c$ and exhibits colossal magnetoresistance (CMR) at T$_c$. To explain all aspects of CMR, a polaron effect due to a strong electron-phonon coupling arising from the Jahn-Teller distortion of the Mn$^{3+}$ ions is necessary [4]. It may be noted that while Mn$^{3+}$ is a Jahn-Teller ion, Mn$^{4+}$ and Mn$^{2+}$ are both non Jahn-Teller ions, and analogous to the hole doped compound (La$_{0.7}$Ca$_{0.3}$MnO$_3$) one could also have an electron doped compound where double exchange could be established between Mn$^{3+}$ and Mn$^{2+}$ [5]. In some prior work done by us and other workers it was shown that if La$^{3+}$ in LaMnO$_3$ be partially substituted by Ce$^{4+}$ ion, thereby driving some of the Mn$^{3+}$ ions to a Mn$^{2+}$ state, it is possible to have double exchange between Mn$^{2+}$ and Mn$^{3+}$ and the compound La$_{0.7}$Ce$_{0.3}$MnO$_3$ undergoes metal insulator transition (MIT) at around 250 K and also exhibits CMR [6,8,9]. Henceforth the hole doped manganite La$_{0.7}$Ca$_{0.3}$MnO$_3$ will be referred to as *p*-manganite and the electron doped manganite La$_{0.7}$Ce$_{0.3}$MnO$_3$ as *n*-manganite.

At room temperature, both the hole doped (*p*-type) and electron doped (*n*-type) manganites are semiconductors as formation of polarons (or local lattice distortions) localize the charge carriers and this localization is lifted only below T$_C$. However they differ from the conventional band semiconductors like Si or Ge, since they are polaronic semiconductors. The splitting of $e_g$ levels of the Mn$^{3+}$ ions, induced by Jahn-Teller distortion, opens up a gap at the Fermi surface. Above T$_C$, the electrical conduction takes place by polaron hopping [10] and below T$_C$ the mobility of the charge carriers increases owing to the parallel alignment of the Mn$^{3+}$/Mn$^{4+}$ moments. This causes the gap to vanish, and at low temperatures the two types of manganites behave as ferromagnetic metals. The existence of a gap at room temperature at the Fermi energy in the *p*-type and *n*-type manganites, with their respective Fermi levels well separated in energy scale, presents us with an opportunity to fabricate *p- n* junction diodes. In this letter, we report for the first time the current-voltage (I-V) characteristics of a *p*-manganite/insulator/*n*-manganite trilayer device with $SrTiO_3$ (STO) as the insulator. Study of I-V characteristics in the current perpendicular to plane (CPP) geometry reveals that the junction indeed behaves as a diode at room temperature (300K). At low temperature however, the metallic nature of the manganite layers suppresses this effect and the I-V characteristics become nearly symmetric. Our results show that it is possible to realize rectifying junctions using the polaronic insulator regime above $T_c$ of the CMR compounds.

The epitaxial films (individual layers) were deposited on a 0.5% Nb-doped $SrTiO_3$ (Nb-STO) conducting substrate by pulsed laser deposition (PLD). Bulk polycrystalline manganite targets for PLD were prepared by solid state reaction route [8]. The films were deposited in an off-axis geometry (where the plume grazes the substrate surface), using a 248 nm KrF excimer laser in an oxygen atmosphere [11]. The substrate was kept at 800° C and an oxygen pressure of 400 mTorr was maintained throughout the deposition. The laser energy density was approximately 3 J/cm$^2$ with a repetition rate of 3 Hz. The deposition chamber had a multi-target holder for insitu growth of the multilayers. After deposition the chamber was vented with high purity oxygen and the substrate cooled down to room temperature. The thickness of the bottom layer ($La_{0.7}Ce_{0.3}MnO_3$) is 500 Å, $SrTiO_3$ (STO) layer is 50 Å, and that of the top layer ($La_{0.7}Ca_{0.3}MnO_3$) is also 500 Å. The thickness of the individual layers was previously calibrated against the number of pulses. Magnetization measurement of the device revealed that both the individual manganite layers had a $T_C$ of 250 K. The room temperature resistance of a 1cm × 1cm substrate when measured across two extreme ends, along the diagonal was found to be ~ 200 Ω.

The I-V characteristics were measured in CPP geometry, such that a current lead and a voltage lead were connected with silver paste at the bottom of the conducting substrate, and the other current and voltage leads were attached to the top $La_{0.7}Ca_{0.3}MnO_3$ layer, such that the electrons (or the holes) have to tunnel through the insulating barrier on their way from one manganite layer to the other. The entire arrangement is shown in figure 1 (inset). The room temperature I-V curve is also shown in figure 1. One can clearly see the asymmetry in the forward and reverse biased mode. This asymmetry reduces upon cooling below $T_C$ and the IV curve at 60K (fig 1) is roughly symmetric.

To understand the above results we first consider the IV characteristic of an ideal tunnel junction. The current in a tunnel junction is given within the WKB approximation as [12]

$$I(s,V,W,T) = A \int_{-\infty}^{\infty} N_1(E+\frac{eV}{2})N_2(E-\frac{eV}{2})T(s,E,W)\{f(E-\frac{eV}{2}) - f(E+\frac{eV}{2})\}dE \quad (1)$$

where $m$ is the electronic mass, $N_1(E)$ and $N_2(E)$ the density of states (DOS) of two metals across the insulating barrier, W is the height of the tunnelling barrier, s its width, $f(E)$ is the Fermi-Dirac function and T(s,E,W) is given by $T(s,E,W) = \exp\{-\frac{m}{\eta}d\sqrt{W-E}\}$. It has to be noted here that though the integration should ideally run from minus infinity to infinity within the WKB approximation it is limited to E<W where T(s,E,W) is real. This however introduces a negligible error if V<<W since the integrant drops rapidly to zero for |E|>V. It is clear from equation (1) that the integral is symmetric with respect to V apart from the contribution arising from the DOS factors $N_1(E)$ and $N_2(E)$. Therefore the asymmetry in a tunnel junction will reflect the asymmetry in the DOS of the two electrodes across the insulating barrier.

To explain the asymmetry at 300K, one has to note that at room temperature the electron and hole doped manganites behave as heavily doped *n* and *p* type semiconductors. In perovskite manganites the $e_g$ band gets split due to Jahn-Teller distortion into two sub-bands, $e_g^1$ and $e_g^2$ separated by a gap of around 0.1eV, the width of each being 1.0 eV [13]. At room temperature (above $T_C$) the Fermi level will lie in the gap opened by polarons. It will be close to the top edge of the $e_g^1$ band in the hole doped manganite, and close to the bottom of the $e_g^2$ band in the electron doped manganite. Thus in the semiconducting regime, the difference in the Fermi level in $La_{0.7}Ca_{0.3}MnO_3$ and $La_{0.7}Ca_{0.3}MnO_3$ with respect to their respective valence and conduction band will ensure that $N_1(E)$ and $N_2(E)$ are different. (It should be noted here that in the case of semiconductors $N_1(E)$ is the density of states of the highest occupied energy state in the semiconductor from which the electron is tunneling, $N_2(E)$ being the density of states of unoccupied state on the other side at the same energy where the electron will tunnel to). The situation is similar to an *n* and *p* type semiconductor separated by barrier and is schematically shown in figure 2. At zero bias $N_1(E)N_2(E)=0$. When a forward bias voltage is applied the product $N_1(E)N_2(E)$ becomes nonzero when the voltage exceeds a threshold value, $V_{th}$. On the other hand when a reverse bias is applied $N_1(E)N_2(E)$ becomes nonzero only when the voltage exceeds $(V_{th}+\Delta/e)$, where $\Delta$ is the bandgap of the semiconductor. As a result the I-V curve mimics the behavior of a diode. However, these features get smeared due to several reasons. Firstly, thermal effect play a significant role since the value of the bandgap is of the order of 0.1eV, which is smaller than typical semiconductors like Si or Ge. Secondly, the leakage current through the non-ideal barrier also contributes in the reverse biased mode. The expected IV curves with and without thermal smearing are shown in figure 2(b). Assuming that the bandgap does not change significantly upon doping, a crude estimate of the bandgap from the I-V characteristics at room temperature yields a value of around 0.45eV which is larger than the activation energy deduced from transport measurements[15]. However, this result clearly shows that a rectifying junctions can be fabricated using hole and electron doped manganites.

The afore mentioned asymmetry in the I-V characteristics of the same device vanishes at low temperature (60K) (fig. 1) and we see a more or less symmetric I-V curve. As mentioned before the polaronic gap disappears below $T_C$ (250 K), and as we go down to very low temperature, the Jahn-Teller split bands, $e_g^1$ and $e_g^2$ merge into one another. This causes the Fermi level to lie well within the conduction band. An asymmetry however can still exist here if the $N_1(E)$ and $N_2(E)$ are very much different in the two compounds [14]. The symmetric I-V curve of the tunnel junction at low temperature essentially implies that the density of states close to Fermi level are similar in their ground states. This is supported by recent band structure calculation in these two compounds [9]. However a careful examination of the data at 60K reveals that a small degree of asymmetry persists even at this temperature implying that small differences in N(E) close to $E=E_F$ exist in these compounds. In order to rule out any possible existence of a Schottky barrier between the manganite and the Nb-STO substrate, we have deposited an *n*-manganite layer on Nb-STO substrate and the I-V curve across the manganite/Nb-STO interface was found to be linear.

In summary we have successfully fabricated an epitaxial trilayer consisting of a hole and an electron doped manganite on either side of an insulating STO barrier, which exhibits strong diode like I-V characteristics at room temperature. We found that the asymmetry in the I-V curve disappears at low temperatures, owing to the insulator to metal transition. The fabrication of a rectifying junction using the polaronic regime of the electron and hole doped compounds might open up new possibilities in the fabrication of electronic devices using these compounds.

CM and PR would like to acknowledge Deutsche Forschungsgemeinschaft (SFB 422) and the Leverhulme Trust respectively for financial support.

**Figure Captions**

Figure 1: The I-V characteristics of the CMR/Insulator/CMR trilayer, measured at room temperature (300K) (●) and at 60K (?) in the CPP mode. Inset shows the schematic diagram of the *p*-manganite/insulator/*n*-manganite trilayer deposited on conducting Nb-STO substrate. The top is the hole doped manganite $La_{0.7}Ca_{0.3}MnO_3$, the bottom layer is the electron doped manganite

La$_{0.7}$Ce$_{0.3}$MnO$_3$ and the insulator being SrTiO$_3$ (STO). The diagram also shows the current and voltage leads.

Figure 2: (a) Schematic diagram of tunnelling across a tunnel junction with *n* and *p* type semiconductors. In forward bias the electron can tunnel through the barrier when V>V$_{th}$. In reverse bias for the electron to tunnel from the *n*-type semiconductor to the *p*-type semiconductor the applied voltage has to be larger than (V$_{th}$+$\Delta$/e). In the diagram the density of states is plotted horizontally vs. the energy vertically. (b) I-V characteristics with (thin line) and without (thick line) thermal smearing.

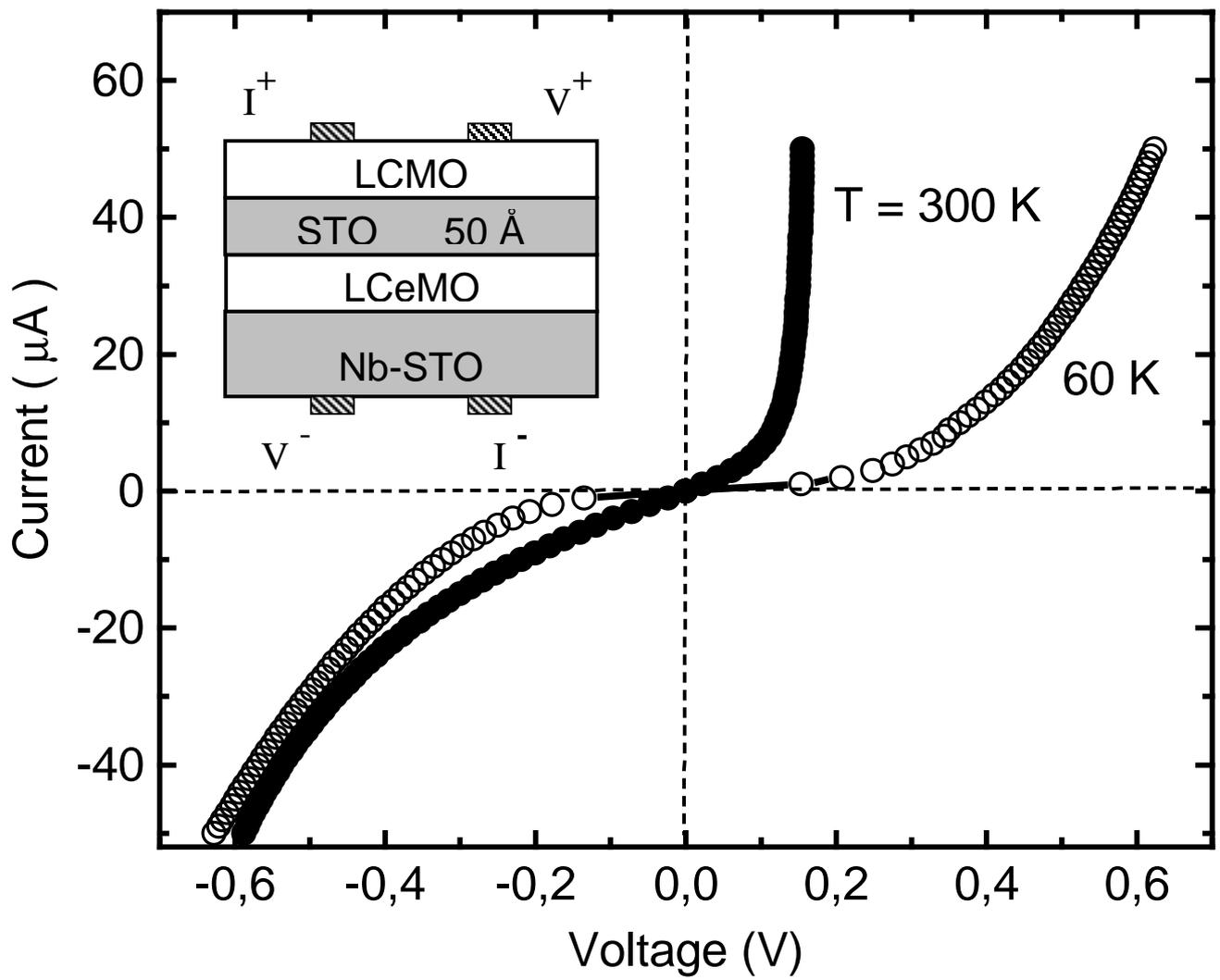

Fig.1

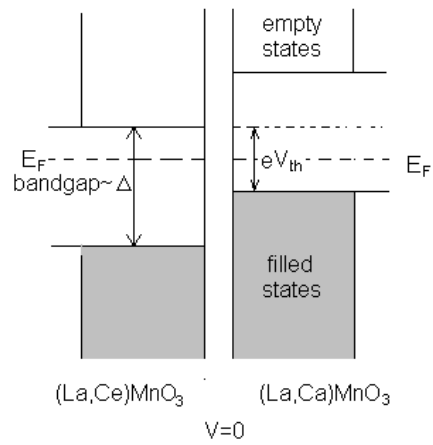

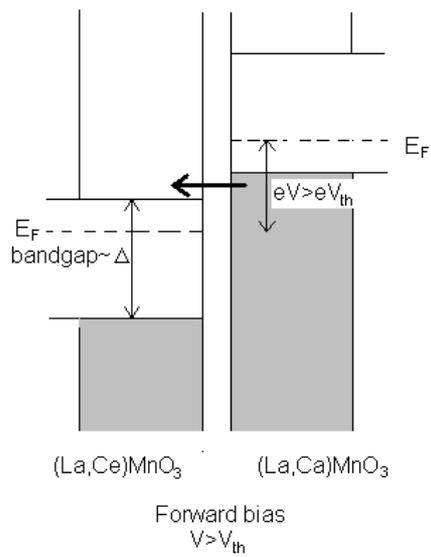

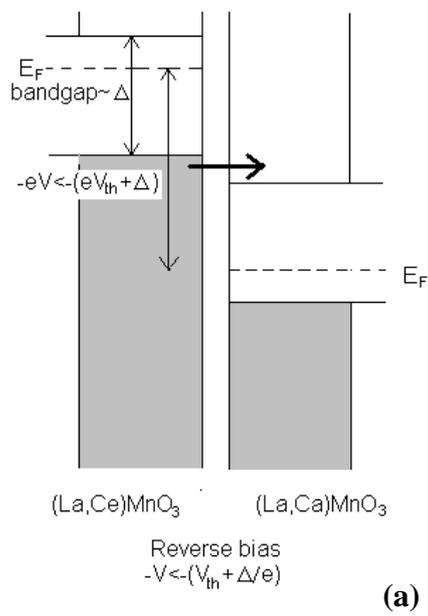

(a)

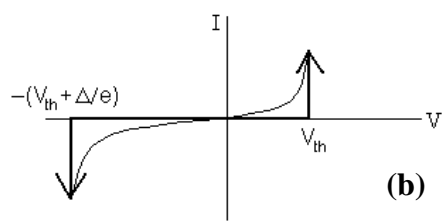

(b)